\documentclass[10pt,conference]{IEEEtran}
\usepackage{cite,graphicx,psfrag,amsmath,amssymb,url,algorithm,algorithmic,color}
\newtheorem{theorem}{Theorem}

\def\Cost{L}
\def\CampCost{L}

\def\boldl{{\mbox{\boldmath $l$}}}

\def\lg{{\log_2}}
\def\boldp{{\mbox{\boldmath $p$}}}

\def\P{{\mathbb P}}

\def\Rp{{\mathbb R}_+}

\def\X{{\mathcal X}}
\def\Z{{\mathbb Z}}

\newcommand{\defn}[0]{\it}

\begin{document}
\title{Reserved-Length Prefix Coding} \bibliographystyle{IEEEtran}

\author{\authorblockN{Michael B. Baer}
\authorblockA{Ocarina Networks\\
42 Airport Parkway\\
San Jose, California  95110-1009  USA\\
Email:{\color{white}{i}}calbear{\color{black}{@}}{\bf \scriptsize \.{1}}eee.org}}

\maketitle

\begin{abstract}
Huffman coding finds an optimal prefix code for a given probability mass function.  Consider situations in which one wishes to find an optimal code with the restriction that all codewords have lengths that lie in a user-specified set of lengths (or, equivalently, no codewords have lengths that lie in a complementary set).  This paper introduces a polynomial-time dynamic programming algorithm that finds optimal codes for this reserved-length prefix coding problem.  This has applications to quickly encoding and decoding lossless codes.  In addition, one modification of the approach solves any quasiarithmetic prefix coding problem, while another finds optimal codes restricted to the set of codes with $g$ codeword lengths for user-specified~$g$ (e.g., $g=2$).
\end{abstract}

\section{Introduction} 
\label{intro} 

A source emits symbols drawn from the alphabet $\X = \{ 1, 2, \ldots,
n \}$.  Symbol $i$ has probability $p_i$, thus defining probability
mass function vector $\boldp$.  We assume without loss of generality that
$p_i > 0$ for every $i \in \X$, and that $p_i \leq p_j$ for every
$i>j$ ($i,j \in \X$).  The source symbols are coded into binary
codewords.  The codeword $c_i$ corresponding to symbol $i$ has length
$l_i$, thus defining length vector $\boldl$.

It is well known that Huffman coding~\cite{Huff} yields a prefix code
minimizing $$\sum_{i \in \X} p_i l_i$$ given the natural coding
constraints: the integer constraint, $l_i \in \Z_+$, and the
Kraft (McMillan) inequality~\cite{McMi}:
\begin{equation}
\kappa(\boldl) \triangleq \sum_{i \in \X} 2^{-l_i} \leq 1 .
\label{kraft}
\end{equation}
Since an exchange argument (e.g., \cite[pp.~124-125]{CoTh2}) easily shows that
an optimal code exists which has monotonic nondecreasing lengths, 
we can assume without loss of generality that such
minimum-redundancy codes have $l_i \geq l_j$ for every $i>j$ ($i,j \in
\X$).

There has been much work on solving this problem with other objectives
and/or additional constraints\cite{Abr01}.  One especially useful
constraint\cite{MoTu97,WMB} is that of length-limited coding, in which
$$l_i \in \{1, 2, \ldots, l_{\max}\} \forall i$$ for some $l_{\max}$.
A constraint that has received less attention is the {\defn
reserved-length} constraint:
$$l_i \in \Lambda = \{\lambda_1, \lambda_2, \ldots ,
\lambda_{|\Lambda|}\} \forall i$$ for $\lambda_i \in \Z_+ \forall i$.
In this case, instead of restricting the range of codeword lengths to
an interval as in length-limited coding, it is restricted to an
arbitrary set of lengths.  (As demonstrated in the next section, there is no loss of generality in assuming this set to be finite).  The problem is well-formed if and only if
$\lambda_{|\Lambda|} \geq \log_2 n$.

This problem was proposed in the 1980s but, due to the lack
of a solution, never published\cite{Zhan}.  
A practical application is that of fast data decompression.
Perhaps the greatest bottleneck in fast Huffman decoding is the
determination of codeword length from input bits, which can be done
using a lookup table, a linear search, or a decision tree,
depending on the complexity of the code involved\cite{MoTu97}.  The
average time taken by a linear search or an optimal decision tree
increases with the number of possible codeword lengths, so limiting
the number of possible codeword lengths can make decoding faster; if the
resulting increase in expected codeword length is small or zero, this can be an
effective way of trading off compression and speed, with no
compression on one end of the spectrum and optimal compression on the
other end.

Consider the optimal prefix code for random variable $Z$ drawn from
the Zipf distribution with $n = 2^{12}$, that is,
$$\P[Z = i]=\frac{1}{i \sum_{j=1}^n j^{-1}}$$
which is approximately equal to
the distribution of the $n$ most common words in the English
language\cite[p.~89]{Zipf}.  This code has codewords of $13$ different
lengths, with an average length of about $8.78$ bits.  If one were to
restrict this code to only allow codewords of lengths in $\{5,9,14\}$,
the resulting optimal restricted code would have an average length of
about $9.27$ bits.  Although suboptimal, this restricted code would
decode more quickly than the optimal unrestricted code.

An $O(n^4)$-time $O(n^3)$-space dynamic programming
approach, introduced shortly, finds optimal reserved-length binary prefix codes.  Variants of this algorithm solve a related length constraint and any
case of the quasiarithmetic coding problem introduced by
Campbell\cite{Camp1}, extending the result of \cite{Baer06}.

\section{Preliminaries and Algorithm} 
\label{alg}

Many prefix coding problems --- most notably binary Huffman coding and
binary length-limited ``Huffman'' coding --- must return an optimal
code in which the Kraft inequality (\ref{kraft}) is satisfied with
equality, that is, for which $\kappa(\boldl) = 1$.  For nonbinary
problems, although the corresponding inequality is not always
satisfied with equality, a simple modification to the problem changes
this, causing the inequality to always be equal for optimal
codes\cite{Huff,BaerI20}. This is not the case for the reserved-length
problem.  For example, if $n=3$ and the allowed lengths are $1$ and
$3$, then the optimal code must have lengths $1$, $3$, and $3$,
resulting in a code for which $\kappa(\boldl) = 0.75$.  Moreover, it
is not clear how to determine $\kappa(\boldl)$ for the optimal code
other than to calculate the optimal code itself.  The Huffman coding
and most common length-limited appoaches rely on $\kappa(\boldl) = 1$,
so these methods cannot be used to find an optimal code here.

The Kraft inequality is often explained in terms of a {\defn coding
tree}.  A binary coding tree is a rooted binary tree in which the
leaves represent items to be coded.  Along the path to a leaf, if the
$j$th edge goes to the leftmost child, the $j$th bit of the codeword
is a~$0$; otherwise, it is a~$1$.  For a finite code tree, the Kraft
inequality is an equality if and only if every node has $0$ or $2$
children, that is, if it is full.  This assumption needs to be relaxed
for finding an optimal reserved-length prefix code.

One approach that does not require $\kappa(\boldl)=1$ is dynamic
programming.  Many prefix coding solutions use dynamic programming
techniques\cite{Abr01}, e.g., finding optimal codes for which all
codewords end with a `1' bit\cite{ChGo}, a situation in which,
necessarily, a finite code cannot have $\kappa(\boldl)=1$.  For the
current problem, the dynamic programming algorithm should find, for
increasing tree heights, a set of candidate trees from which to
choose, and it should terminate when the longest feasible length is
encountered.  First, however, we have to find this longest feasible
length, since we didn't specify that $\Lambda$, the set of allowed
lengths, needed to be upper-bounded by any function of $n$ or even
finite.

\begin{theorem}
Any codeword $l_i$ of an optimal reserved-length code either satisfies
$l_i \leq n-2$ or $l_i=\lambda_{\infty}$, where $\lambda_{\infty}$ is
the smallest element of $\Lambda$ that satisfies $\lambda_{\infty} >
n-2$.
\label{lambda}
\end{theorem}

\begin{proof}
We first show that no partial Kraft sum of $x$ items
$$\kappa(\boldl,x) \triangleq \sum_{i=1}^x 2^{-l_i}$$ can be in the
open interval $(1-2^{-x},1)$, and, furthermore, if the longest
codeword is of length $l_x>x-1$, the sum cannot be in
$(1-2^{-x+1}+2^{-l_x},1)$.  This is shown by induction on codeword
lengths of nondecreasing order.  Clearly
$$\kappa(\boldl,2) = 2^{-l_1} + 2^{-l_2} \notin (3/4, 1)$$ satisfies
this.  Suppose the Kraft sum for $x-1$ items cannot fall in
$(1-2^{-x+1},1)$, that is, for any code for which
$\kappa(\boldl,x-1)<1$, $\kappa(\boldl,x-1) \leq 1-2^{-x+1}$.  Since
the $x$th term is a power of two, the partial sum of a code is no
greater than $1-2^{-x+1}+2^{-x} = 1-2^{-x}$ for $\kappa(\boldl,x)<1$.
Moreover, if $l_x \geq x$, the partial sum is less than or equal to
$1-2^{-x+1}+2^{-l_x}$.

Now suppose there is an optimal code for $n$ items which includes
codeword lengths $l_\mu$ and $l_\nu$, where $n-2<l_\mu<l_\nu$.  Assume
without loss of generality that $l_\mu$ and $l_\nu$ are the longest
codeword lengths and $l_\nu=l_n$ (i.e., $l_\nu$ is the longest
codeword length).  Note that $l_\nu \geq n$ and the Kraft sum cannot
equal $1$ for any code in which the longest codeword has length equal
to or exceeding~$n$; it is well known that the deepest full tree is a
terminated unary tree, one with depth $n-1$.  Thus $\kappa(\boldl) <
1-2^{-n}$.  Consider a code with lengths $l'_i = l_i$ for $i<n$ and
$l'_n = l_\mu$.  We show that a prefix code exists with these lengths
and thus achieves greater compression, rendering $\boldl$ suboptimal.
If $l_\nu = l_\mu + 1$, then $$\kappa(\boldl') = \kappa(\boldl) -
2^{-l_\mu-1} + 2^{-l_\mu} \leq 1-2^{-n}+2^{-l_\mu-1} \leq 1$$ since $n
\leq l_\mu+1$.  Otherwise, $l_n=l_\nu \geq n$, and $$\kappa(\boldl') =
\kappa(\boldl) - 2^{-l_\nu} + 2^{-l_\mu} \leq 1-2^{-n+1}+2^{-l_\mu}
\leq 1$$ since $n-1 \leq l_\mu$.
\end{proof}

Since an optimal tree exists which has monotonic nondecreasing
lengths, optimal codeword lengths can be fully specified by the number
of leaves on each of the allowed levels of the code tree.  For such an
optimal tree, given any ``allowed level'' $\lambda_m$, the lengths
with $l_i \leq \lambda_m$ have a partial Kraft sum
$$\kappa_{\lambda_m}(\boldl) \triangleq \kappa(\boldl,\upsilon_m)$$
for $\upsilon_m$ such that $l_{\upsilon_m} \leq \lambda_m$ and either
$l_{1+\upsilon_m} \leq \lambda_m$ or $\upsilon_m = n$.  This Kraft sum
is a multiple of $2^{-\lambda_m}$, so there exists an $\eta_m$ such
that $\kappa(\boldl,\upsilon_m) = 1-\eta_m2^{-\lambda_m}$, and this
$\eta_m$ is the number of internal nodes on level $\lambda_m$ of any
coding tree corresponding to the codeword lengths.

In an optimal coding tree, if $\Delta_m$ is defined to be
$\lambda_{m+1} - \lambda_{m}$, then, for any $\upsilon_m < n$,
\begin{equation}
\underbrace{\eta_m2^{\Delta_m} - (2^{\Delta_m} - 2)}_{\begin{array}{c}
\mbox{internal nodes next minus}\\
\mbox{single-node expansion factor}
\end{array}} \leq \underbrace{n-\upsilon_m}_{\mbox{leaves under $m$}}.
\label{eta1}
\end{equation}
This can be seen by observing that, if a code violates
this, we can produce a code with the same lengths for $l_1$ through~$l_{\upsilon_m}$ and assign $l_{\upsilon_m+1} = \lambda_m$ and $l_i =
\lambda_{m+1}$ for $i>\upsilon_m+1$, and the new code would have no
length exceeding that of the original code; in fact,
$l_{\upsilon_m+1}$ is strictly shorter, so the original code could not
be optimal.  For $\lambda_{m+1} = \lambda_{m}+1$, this condition is
identical to 
\begin{equation}
2\eta_m \leq n-\upsilon_m
\label{eta2}
\end{equation}
which is a looser necessary condition for optimality.  For similar
reasons, no optimal tree will have a partial tree with
$\upsilon_m=n-1$ for any~$m$, since using an internal node on level
$\lambda_m$ for the final item results in an improved tree.

Such properties can be used to construct a dynamic programming algorithm.  In describing this algorithm, we use the following notation (with mnemonics in boldface):

$$
\begin{array}{ll}
\upsilon_m:&\mbox{ \textbf{U}sed u\textbf{p} leave\textbf{s} at or above level } \lambda_m \\
\eta_m:&\mbox{ Nod\textbf{e}s in\textbf{t}ern\textbf{a}l at level } \lambda_m \\
\Upsilon[m,\upsilon,\eta]:&\mbox{ Leaves above level } \lambda_m \\
\Cost[m,\upsilon,\eta]:&\,\,\sum_{i=1}^{\upsilon_m}p_il_i+\lambda_m \sum_{i=1+\upsilon_m}^n p_i \\
\end{array}
$$

The idea for the algorithm is to calculate the optimal
$\Cost[m,\upsilon_m,\eta_m]$ given feasible values of partial trees
($\upsilon_m < n-1$) and to separately keep track of the best finished
tree ($\upsilon_m = n$) as the algorithm progresses.  The trees grow
by level ($\lambda_m$ for increasing~$m$), while the algorithm
calculates all feasible values of $\upsilon_m$ (which are in $[0,n-2]$
for partial trees) and $\eta_m$ (which are in $[0,\lfloor n/2
\rfloor]$ for partial trees due to (\ref{eta2}); if $\eta_m > n/2$, at
least one node on a lower level could be shortened to length
$\lambda_m$, resulting in a strictly improved code).  Thus there are
$O(n^2)$ values per level, and we can try all feasible combinations,
calculating $\Cost$ for all combinations of partial trees --- saving
optimal combinations --- and finished trees --- saving only the best
finished tree encountered up to this point.  Clearly, $\upsilon_m$
must be nondecreasing.  This, along with the bounds on $\eta_m$, are
used to try the aforementioned combinations.  In cases where
$|\Lambda|$ is much smaller than~$n$, additional constraints can be
made, based on (\ref{eta1}), but such constraints do not improve
computational complexity in the general case, so we do not discuss
them here.

After finishing level $\lambda_{|\Lambda|}$, the optimal tree is
rebuilt via backtracking.  Assuming arithmetic operations are
constant-time, complexity of the dynamic programming
Algorithm~\ref{dpfig} is
$O(|\Lambda|n^3)$-time and $O(|\Lambda|n^2)$-space.
Because $|\Lambda|<n$ without loss
of generality, if we assume arithmetic operations are constant time, time
complexity should be $O(n^4)$ and space complexity $O(n^3)$.

\begin{algorithm*}
\caption{Dynamic programming algorithm for reserved-length prefix coding}\label{program}
\begin{algorithmic}[1]
\REQUIRE $\boldp$ of size $n=|\X|$, $\Lambda$ for which (without loss of generality) $\lambda_{|\Lambda|-1} 
\leq |\X| - 2$
  \STATE $F_0 \leftarrow 0$ 
  \FOR{$i \leftarrow 1,|\X|$}
    \STATE $F_i \leftarrow F_{i-1} + p_i$ \COMMENT{Calculate cumulative distribution function}
  \ENDFOR

  \FORALL{$0 \le m < |\Lambda|$, $0 \le \upsilon \leq |\X|-2$, $0 \le \eta \le \lfloor |\X|/2 \rfloor$} 
    \STATE $\Cost[m,\upsilon,\eta] \leftarrow \infty$ \COMMENT{Initialize partial tree costs}
  \ENDFOR
  \STATE $\Cost_{\min} \leftarrow \infty$ \COMMENT{Best total tree cost so far}
  \STATE $\Cost[0,0,1] \leftarrow 0$ \COMMENT{Trivial tree cost}
  \STATE $\lambda'' \leftarrow 0$ \COMMENT{Previous level}
  \FOR{$m \leftarrow 1,|\Lambda|$ \COMMENT{Level by level}}
    \STATE $\lambda' \leftarrow \lambda_m$ \COMMENT{Current level}
    \FORALL{$(\upsilon,\eta) \in [0,|\X|-2] \times [0, \lfloor |\X|/2 \rfloor]$ \COMMENT{Find optimal partial trees with given $m$ from $(m-1,\upsilon,\eta)$}}
      \IF{$\Cost[m-1,\upsilon,\eta] < \infty$}
         \STATE $\eta' \leftarrow \eta \ll (\lambda'-\lambda'')$ \COMMENT{Total nodes on new level $\lambda_m$}         \STATE $\Cost' \leftarrow \Cost[m-1,\upsilon,\eta] + (\lambda'-\lambda'')(1-F_\upsilon)$ \COMMENT{Cost on new level $\lambda_m$}         \IF{$m < |\Lambda|$ \COMMENT{Build partial trees (for which $m<|\Lambda|$)}}
           \STATE $\upsilon_{\min} \leftarrow \max(\upsilon,2(\upsilon+\eta')-|\X|)$ \COMMENT{Range of potential $\upsilon_m$}
           \STATE $\upsilon_{\max} \leftarrow \min(\upsilon + \eta',|\X|-2)$
           \FOR{$\upsilon' \leftarrow \upsilon_{\min},\upsilon_{\max}$ \COMMENT{Compare cost for all potential $\upsilon_m<|\X|$}}
             \IF{$\Cost[m,\upsilon',\eta'-\upsilon'+\upsilon] > \Cost'$}
               \STATE $\Cost[m,\upsilon',\eta'-\upsilon'+\upsilon] \leftarrow \Cost'$ \COMMENT{New optimal partial cost for $(m, \upsilon_m \eta_m) = (m, \upsilon', \eta'-\upsilon'+\upsilon)$}
               \STATE $\Upsilon[m,\upsilon',\eta'-\upsilon'+\upsilon] \leftarrow \upsilon$ \COMMENT{Save with $\upsilon_{m-1}$ for backtracking}
             \ENDIF
           \ENDFOR
         \ENDIF
      \ENDIF
      \IF{$|\X| \leq \upsilon+\eta'$}
         \IF{$\Cost' < \Cost_{\min}$ \COMMENT{Find best finished tree}}
               \STATE $\Cost_{\min} \leftarrow \Cost'$ \COMMENT{Best finished tree cost}
               \STATE $(m_{\min},\upsilon_{\min},\eta_{\min}, \chi_{\min}) \leftarrow (m, |\X|, \eta'-|\X|+\upsilon, \upsilon)$ \COMMENT{Save optimal values with $\chi_{\min}=\upsilon_{m-1}$ for backtracking}
         \ENDIF
      \ENDIF
    \ENDFOR
    \STATE $\lambda'' \leftarrow \lambda'$ \COMMENT{Current level now previous level}
  \ENDFOR

  \STATE $(m,\upsilon,\eta,\chi) \leftarrow (m_{\min},\upsilon_{\min},\eta_{\min},\chi_{\min})$ \COMMENT{Backtrack to find optimal tree}
  \STATE $c \leftarrow (1 \ll \lambda_m) - \eta$ \COMMENT{$1$ greater than integer representation of final codeword}
  \WHILE{$m > 1$ \COMMENT{Rebuild best tree}}
    \IF{$\upsilon < |\X|$}
      \STATE $\chi \leftarrow \upsilon-\Upsilon[m,\upsilon,\eta]$ \COMMENT{Number of leaves above level}
    \ENDIF
    \FOR{$j \leftarrow \upsilon$ down to $\upsilon-\chi+1$}
      \STATE $(l_j, c_j) \leftarrow (\lambda_m, \{c+j-\upsilon-1\}_{\lambda_m})$ \COMMENT{Assign lengths/codewords (where $\{x\}_y$ denotes the $y$-bit representation of $x$)}
    \ENDFOR
    \STATE $c \leftarrow c \gg (\lambda_m - \lambda_{m-1})$ \COMMENT{Start codewords of length $\lambda_{m-1}$}
    \STATE $(m,\upsilon,\eta) \leftarrow (m-1,\upsilon - \chi,(\eta + \chi) \gg (\lambda_m - \lambda_{m-1}))$ \COMMENT{Calculate new $(m,\upsilon,\eta)$ from old using $\chi$}  \ENDWHILE
  \FOR{$j \leftarrow \upsilon$ down to $1$}
     \STATE $(l_j, c_j) \leftarrow (\lambda_1, \{j-1\}_{\lambda_1})$ \COMMENT{Shortest lengths/codewords}
  \ENDFOR
\label{dpfig}
\end{algorithmic}
\end{algorithm*}

A simple example of this algorithm at work is in finding an optimal
code for Benford's law\cite{Newc,Benf} with the restriction that all
codeword lengths must be powers of two.  In this case, $p_i$ is
$\log_{10}(i+1) - \log_{10}(i)$ for $i$ from $1$ to $n=|\X|=9$, and
$\Lambda = \{1,2,4,8\}$ is a sufficient range of lengths to allow, due
to Theorem~\ref{lambda}.  The calculated values for each feasible partial
$\Cost[m,\upsilon,\eta]$ are shown in Table~\ref{example}.

On the first level, $\lambda_1$, average length is identical to the
level number, and, if, for example, $\lambda_1=1$, the nodes at the
level can include zero ($(\upsilon,\eta)=(0,2)$), one ($(1,1)$), or
two ($(2,0)$) terminating nodes, which are the only nontrivial entries
in a two-dimensional grid for this
level, as indicated by the first grid in Table~\ref{example}.  
From each nontrivial entry in the level $\lambda_1$ grid, all allowed
combinations of terminating and expanding are considered until the
second (level $\lambda_2$) grid is arrived at, and the algorithm
proceeds similarly until all allowed levels are accounted for.  All
trees with $\upsilon=n$ (all leaves accounted for) are compared with
the best one so far in order to find an optimal tree.  In the
Benford's law example, this is a tree with two codewords of length two
and seven codewords of length four.  Note that the strict inequality
of line 29 means that, if there are multiple optimal length vectors,
the algorithm selects one of minimal maximum length.

\begin{table*}[t]
$$
\begin{array}{|r||rrrrrrrr|}
\multicolumn{9}{c}{p_{\mbox{\tiny Benford}} \approx 
\{0.301, 0.176, 0.125, 0.097, 0.079, 0.067, 0.058, 0.051, 0.046\},
\quad \Lambda=(1,2,4,8)} \\[4pt]
\multicolumn{9}{c}{\mbox{Level } \lambda_1 = 1 \quad (m=1)} \\
\hline
&\upsilon_1=0&1&2&3&4&5&6&7\\
\hline
\hline
\eta_1=0&\infty&\infty&1.000 (0)&\infty&\infty&\infty&\infty&\infty\\
1&\infty&1.000 (0)&\infty&\infty&\infty&\infty&\infty&\infty\\
2&\textbf{1.000 (0)}&\infty&\infty&\infty&\infty&\infty&\infty&\infty\\
3&\infty&\infty&\infty&\infty&\infty&\infty&\infty&\infty\\
4&\infty&\infty&\infty&\infty&\infty&\infty&\infty&\infty\\
\hline
\multicolumn{9}{c}{} \\
\multicolumn{9}{c}{\mbox{Level } \lambda_2 = 2 \quad (m=2)} \\
\hline
&\upsilon_2=0&1&2&3&4&5&6&7\\
\hline
\hline
\eta_2=0&\infty&\infty&1.523 (2)&1.699 (1)&2.000 (0)&\infty&\infty&\infty\\
1&\infty&\infty&1.699 (1)&2.000 (0)&\infty&\infty&\infty&\infty\\
2&\infty&1.699 (1)&\textbf{2.000 (0)}&\infty&\infty&\infty&\infty&\infty\\
3&\infty&2.000 (0)&\infty&\infty&\infty&\infty&\infty&\infty\\
4&2.000 (0)&\infty&\infty&\infty&\infty&\infty&\infty&\infty\\
\hline
\multicolumn{9}{c}{} \\
\multicolumn{9}{c}{\mbox{Level } \lambda_3 = 4 \quad (m=3)} \\
\hline
&\upsilon_3=0&1&2&3&4&5&6&7\\
\hline
\hline
\upsilon_3=0&\infty&\infty&2.569 (2)&2.495 (3)&2.602 (4)&\infty&2.745 (2)&2.796 (3)\\
1&\infty&\infty&\infty&\infty&\infty&2.745 (2)&2.796 (3)&\infty\\
2&\infty&\infty&\infty&\infty&2.745 (2)&2.796 (3)&\infty&\infty\\
3&\infty&\infty&\infty&2.745 (2)&\infty&\infty&\infty&\infty\\
4&\infty&\infty&\infty&\infty&\infty&\infty&\infty&\infty\\
\hline
\end{array}
$$
\caption{Grids for finding reserved-length solution via
dynamic programming.  Each value represents optimal partial cost for a
given $\eta$, $\upsilon$, and level $\lambda$, with number of leaves
above the given level given in parentheses.  The partial trees used in
the optimal result --- that terminated with $(m, \upsilon, \eta) = (3, 9,
2)$ having $\boldl = \{2, 2, 4, 4, 4, 4, 4, 4, 4\}$ --- are shown in
boldface.}
\label{example}
\end{table*}

Note that a similar approach could be used for nonbinary trees,
although an efficient exponentiation procedure should be used in place
of shifting in lines 15 and 47 of Algorithm~\ref{dpfig}.  Codeword
construction changes (lines with ``$c_j$'') and the aforementioned
expansion bounds (lines with ``$|\X|/2$'') also need adjustment for
nonbinary cases.  These alternations do not worsen computational
complexity.

\section{Extensions and Conclusion}

The aforementioned method yields a prefix code minimizing expected
length for a known finite probability mass function under the given
constraints.  However, there are many varied instances in which
expected length is not the proper value to minimize\cite{Abr01}.  
Many such problems are in a certain family of generalizations of the
Huffman problem introduced by Campbell in \cite{Camp}.

While Huffman coding minimizes $\sum_{i \in \X} p_i l_i$, Campbell's
quasiarithmetic formulation adds a continuous (strictly) monotonic
increasing {\defn cost function} $\varphi(l):\Rp \rightarrow \Rp$.
The value to minimize is then
$$
\CampCost(\boldp,\boldl,\varphi) \triangleq \varphi^{-1}\left(\sum_{i \in \X}
p_i \varphi(l_i)\right). $$

Convex $\varphi$ have been solved for \cite{Baer06}.  For nonconvex
functions, it suffices to replace line~16 in the algorithm,
$$\Cost[i-1,\upsilon,\eta] + (\lambda'-\lambda'')(1-F_\upsilon)$$
with
$$\Cost[i-1,\upsilon,\eta] +
(\varphi(\lambda')-\varphi(\lambda''))(1-F_\upsilon).$$ The exchange
argument still holds, resulting in a monotonic solution, and $\Lambda$
still has cardinality less than~$n$, so the algorithm proceeds
similarly for identical reasons, and thus with the same complexity.  A
nonbinary coding extension is similar to that used to minimize
expected length.

We earlier stated that one purpose for reserving lengths is to allow
faster decoding by having fewer codewords.  However, if this is the
objective, the problem remains of how to select the codeword lengths
to use.  We might, for example, restrict our solution to having two
codeword lengths, but not put any restrictions on what these codeword
lengths should be.  Such a problem was examined analytically in
\cite{FiHo} for $n$ approaching infinity.  Here, we consider solving
the problem for fixed~$n$.

One approach to the two-length problem would be to try all feasible
combinations of codeword lengths.  We then have to find a feasible
set, hopefully one relatively small so as not to drastically increase
the complexity of the problem.

First note that, if only one codeword length is used, then
$\lambda_2=\lambda_1=\lceil \lg n \rceil$.  Otherwise, we begin by
observing that, for the best tree, the number of internal nodes and
leaves on the first allowed level $\lambda_1$ must each be greater
than $0$ (or else only one codeword length could be used) and combined
be no greater than $n-1$ (or else a better code exists with all
codewords having one length).  Thus $\lambda_1 \leq \lg (n-1)$, or,
put another way, $\lambda_1 \leq \lceil \lg n \rceil - 1$.  At the
same time, the second allowed level cannot have $2n-2$ or more
combined internal nodes and leaves; otherwise an improved tree can be
found by decreasing $\lambda_2$ by one, since no more than $n-1$
leaves can be on this level.  Because these nodes are all descendants
of all least one internal node on the first allowed level, this
results in $2^{\lambda_2-\lambda_1} < 2n-2$, which leads to
$\lambda_2-\lambda_1 \leq \lceil \lg (n-1) \rceil \leq \lceil \lg n
\rceil$.  Combining these results, we find that $\lambda_2 \leq 2
\lceil \lg n \rceil - 1$.

This result, while not the strictest bound possible, is sufficient for
us to determine that the number of codeword length combinations one
would have to try would be $O(\log^2 n)$.  Thus, since $|\Lambda|=2$
in all cases and only $O(1)$ data need be kept between combinations,
the algorithm has only an $O(n^2)$ space and an $O(n^3 \log^2 n)$ time
requirement, smaller than even the general version of the reserved
length problem.  For example, the optimal two-length code for the
Benford distribution has two codewords of length two and seven
codewords of length four.  This is the code found above to be optimal
for lengths restricted to powers of two.  This two-length code has
average codeword length $3.04\ldots$, very near to that of the optimal
unrestricted Huffman code, which has average codeword length
$2.92\ldots$.

The two-length problem's solution can be easily generalized to that of
a $g$-length problem, which can be optimally solved with $O(n^2 g)$
space and $O(n^3 (\log^g n) g^3)$ time in similar fashion.  In fact,
all $g'$-length problems, for $g' \leq g$, can be solved with this
complexity, allowing for a selection of the desired trade-off between
number of codeword lengths (speed) and expected codeword length
(compression efficiency).  Modifications can enact additional
restrictions on codeword lengths (e.g., a limit on maximum length) in
a straightforward fashion.

We thus find that this dynamic programming method is quite general,
solving three problems that previously had no proposed polynomial-time
solutions: the reserved-length problem, Campbell's quasiarithmetic
problem, and the $g$-length problem.

\ifx \cyr \undefined \let \cyr = \relax \fi

\end{document}